\documentclass[pra,aps,reprint,amsmath,superscriptaddress]{revtex4-1}
\usepackage{graphicx}
\usepackage{dsfont}

\newcommand{\bra}[1]{\langle #1 | \,}
\newcommand{\ket}[1]{\, | #1 \rangle}
\newcommand{\braket}[2]{\langle #1 | #2 \rangle}
\newcommand{\lra}{\leftrightarrow}
\newcommand{\Ga}{\Gamma}
\newcommand{\la}{\lambda}
\newcommand{\de}{\delta}
\newcommand{\De}{\Delta}
\newcommand{\veps}{\varepsilon}
\newcommand{\om}{\omega}
\newcommand{\Om}{\Omega}
\newcommand{\rc}{\mathrm{c}}
\newcommand{\rt}{\mathrm{t}}

\newcommand{\hlf}{\mbox{$\frac{1}{2}$}}

\begin{document}

\title{High-fidelity Rydberg quantum gate via a two-atom dark state}

\author{David Petrosyan}
\affiliation{Department of Physics and Astronomy, Aarhus University,
DK-8000 Aarhus C, Denmark}
\affiliation{Institute of Electronic Structure and Laser, FORTH,
GR-71110 Heraklion, Crete, Greece}

%\author{Andrew C. J. Wade}
%\affiliation{Department of Physics and Astronomy, Aarhus University,
%DK-8000 Aarhus C, Denmark}

\author{Felix Motzoi}
\affiliation{Department of Physics and Astronomy, Aarhus University,
DK-8000 Aarhus C, Denmark}

\author{Mark Saffman}
\affiliation{Department of Physics, University of Wisconsin-Madison,
Madison, Wisconsin 53706, USA}

\author{Klaus M\o lmer}
\affiliation{Department of Physics and Astronomy, Aarhus University,
DK-8000 Aarhus C, Denmark}

\date{\today}

\begin{abstract}

We propose a two-qubit gate for neutral atoms in which one of the logical state components adiabatically follows a two-atom dark state formed by the laser coupling to a Rydberg state and a strong, resonant dipole-dipole exchange interaction between two Rydberg excited atoms.
Our gate exhibits optimal scaling of the intrinsic error probability $E \propto (B\tau)^{-1}$ with the interatomic interaction strength $B$ and the Rydberg state lifetime $\tau$. Moreover, the gate is resilient to variations in the interaction strength, and even for finite probability of double Rydberg excitation, the gate does not excite atomic motion and experiences no decoherence
due to internal-translational entanglement.
\end{abstract}

%\pacs{32.80.Ee, %Rydberg states
%32.80.Rm, %Multiphoton ionization and excitation to highly excited states
%32.80.Qk  %Coherent control of atomic interactions with photons
%03.67.Lx, %Quantum computation architectures and implementations
%}

\maketitle

\section{Introduction}

Strong, long-range interactions between atoms excited to the Rydberg
states with large principal quantum numbers $n$ make them attractive
systems for the studies of few- and many-body physics and for quantum
information applications \cite{rydQIrev}.
Different schemes of interatomic interactions are employed in this research,
ranging from the blockade of multiple Rydberg excitations of nearby atoms
by resonant laser fields \cite{rydQIrev,Lukin2001,Comparat2010},
and the anti-blockade non-resonant (facilitated) laser excitation
\cite{Gaerttner2013,Lesanovsky2014,Schempp2014,Malossi2014,Urvoy2015}, to
the Rydberg dressing of the ground state atoms by very far-off-resonant lasers
\cite{Bouchoule2002,Johnson2010,Henkel2010,Pupillo2010,Zeiher2016,Buchmann2017,Zeiher2017}.

The Rydberg-state interatomic interactions hold unique potential  
for the implementation of quantum gates with spatially separated neutral
atoms. In the seminal proposal of
Jaksch \textit{et al.} \cite{Jaksch2000}, each qubit is encoded
in a pair of (meta-)stable states $\ket{0}$ and $\ket{1}$ of an atom, and
two-qubit gate operations are performed by selectively exciting
a pair of atoms from specific qubit states, e.g. $\ket{1}$,
to the interacting Rydberg states.
In the regime of a weak dispersive interaction, the pair of Rydberg-excited
atoms acquires an interaction-induced phase shift, which is then transferred
to the corresponding two-qubit state $\ket{11}$ by coherently de-exciting 
the atoms. In the alternative regime of strong interaction,
if one (control) atom is resonantly excited to the Rydberg state,
the interaction-induced level shift suppresses Rydberg excitation
of the second (target) atom within a distance of several micrometers.
Ideally, this blockade effect \cite{rydQIrev,Lukin2001,Comparat2010}
does not depend on the precise value of the interaction strength,
as long as it is sufficiently strong to completely preclude multiple
Rydberg excitations. Since at most one atom is excited to the
Rydberg state at a time, the interaction potential does not
induce interatomic forces, which would otherwise entangle the
internal (qubit) and external (motional) degrees of freedom of the atoms.
The Rydberg blockade gate has therefore been the preferred choice
for quantum logic gate operations
\cite{Jaksch2000,NatPRLSaffman,NatPRLGrangier,Beguin2013,Maller2015}.

The performance of the Rydberg blockade gate has been extensively
analyzed \cite{Saffman2005,LIMSKM2011,Zhang2012,Theis2016}, taking into account
various experimental imperfections and fundamental limitations
of the scheme. Assuming that technical errors due to, e.g., laser phase
and amplitude fluctuations and finite temperature atomic motion and
Doppler shifts, can be eliminated, and that 
leakage errors to the unwanted Rydberg states can be suppressed
by using, e.g., shaped laser pulses \cite{Theis2016},
the remaining limitations of the standard blockade
gate stem from the finite lifetime $\tau = 1/\Ga \propto n^3$
of the Rydberg states, duration $T \simeq 2 \pi/\Om$ of the gate
performed by excitation lasers with Rabi frequency $\Om \gg \Ga$,
and finite Rydberg-Rydberg interaction strength $B \gg \Om$.
Two types of intrinsic errors have been identified:
the error $E_\mathrm{decay} \simeq 2 \pi \Ga /\Om$ due to
the decay of the Rydberg states during the gate time $T$, and
the rotation error $E_\mathrm{rot} \simeq \Om^2/2 B^2$ due to
imperfect blockade of double Rydberg excitation.
Minimizing the total error $E =E_\mathrm{decay} + E_\mathrm{rot}$
with respect to $\Om$ leads then to $E \propto (\Ga/B)^{2/3}$
scaling of the intrinsic error \cite{Zhang2012}.
The resulting estimates for the gate error probability
($E \sim 10^{-3}$) are above the required threshold
values for fault tolerant quantum computation.
We recall that scaling the quantum hardware in order to tackle problems
for which quantum computers may outperform their classical counterparts
entails low threshold values of the gate error probabilities:
$E \simeq 2 \times 10^{-5}$ for use of the $[[7,1,3]]$ Steane and 
$[[9,1,3]]$ Bacon-Shor error correction codes \cite{Svore2007,Spedalieri2009},
and $E\simeq 4  \times 10^{-4}$ for use of the Knill C4/C6 code 
\cite{Lai2014}.

Here we propose and analyze an improved mechanism for implementing
entangling two-atom Rydberg gates. Our gate is similar
to the Rydberg blockade gate, but with an important difference.
As in the standard blockade gate, we excite the control and target
atoms from the qubit states $\ket{1_{\rc , \rt}}$ to the Rydberg
states $\ket{r_{\rc , \rt}}$. But instead of relying on the interaction-induced
level shift of a pair of Rydberg states, we employ adiabatic following
of the two-atom dark state that exists when the atoms in state
$\ket{r_\rc r_\rt}$ can undergo strong resonant dipole-dipole exchange
interaction with another Rydberg-product state $\ket{a_\rc b_\rt }$.
By using a smooth $2\pi$ laser pulse to resonantly drive the target atom,
we ensure that all of the residual Rydberg-state population adiabatically
returns back to the qubit state $\ket{1_\rt}$, eliminating thereby the
rotation error $E_\mathrm{rot}$. The minimal intrinsic error then scales
as $E \propto (\Ga/B)$, with $\Om/B \ll 1$, and it can reach values as small
as $E \sim 10^{-5}$ for the Rydberg states with $n \gtrsim 100$.

In the following sections, we present the quantitative description
of our scheme, estimates of the smallest achievable intrinsic error
probability of the gate and the results of numerical simulations.
In Appendix~\ref{ap:RSI} we outline the essentials of the resonant
F\"orster process for properly tuned Rydberg states of atoms to
realize our two-atom dark state, while detailed description of the
two-atom systems is given in Appendix~\ref{ap:Ham}.
In Appendix~\ref{ap:RBG} we show that using adiabatic pulses 
for the conventional blockade gate also eliminates rotation 
errors and improves its performance.
In contrast to the conventional gates, however, in our scheme
the uncertainty in the interaction strength does not lead to phase
errors, and despite the non-vanishing probability of double Rydberg excitation
there is no mechanical force between the atoms, which would otherwise
hinder the operation of both the interaction and blockade gates with
the Rydberg-excited atoms \cite{Li2013,DPKM2014,Rao2014}.

\section{The dark state adiabatic gate}
\label{sec:tads}

%%%%%%%%%%%%%%%%FIGURE%%%%%%%%%%%%%%%%
\begin{figure}[t]
\includegraphics[width=5cm]{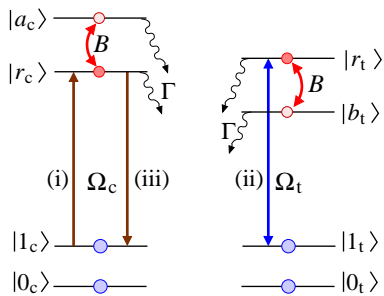}
\caption{(Color online)
Level scheme of two atoms leading to a Rydberg phase
gate by three focused laser pulses.
States $\ket{0_{\rc,\rt}}$ and $\ket{1_{\rc,\rt}}$ are long-lived qubit
basis states of the control ($\rc$) and target ($\rt$) atoms,
and states $\ket{r_{\rc,\rt}}$ and $\ket{a_{\rc}}$, $\ket{b_{\rt}}$ are
Rydberg states with decay rate $\Ga$.
In steps (i) and (iii) the control atom in state $\ket{1_\rc}$
is resonantly excited and de-exited to state $\ket{r_\rc}$ by a
laser with Rabi frequency $\Om_\mathrm{c}$ and pulse area $\pi$, and
in step (ii) the target atom in state $\ket{1_\rt}$ is resonantly
coupled to state $\ket{r_\rt}$ by another laser with Rabi frequency
$\Om_\mathrm{t}$ and pulse area $2\pi$.
Atoms excited to Rydberg states $\ket{r_\rc}$ and $\ket{r_\rt}$ strongly
interact with each other, $B \gg \Om_\mathrm{t}$, via resonant dipole-dipole
exchange process $\ket{r_\rc r_\rt} \lra \ket{a_\rc b_\rt}$ which leads
to suppression of excitation of the target atom in step (ii) if the control
was initially in state $\ket{1_\rc}$.
We assume that atoms in state $\ket{0}$ are decoupled from the laser fields.}
\label{fig:als}
\end{figure}
%%%%%%%%%%%%%%%%%%%%%%%%%%%%%%%%

In Fig.~\ref{fig:als} we show the relevant energy levels of two atoms 
for realizing the Rydberg quantum gate. The qubit basis states are
represented by a pair of long-lived hyperfine ground state sublevels
$\ket{0}$ and $\ket{1}$ which can be manipulated by a
microwave (MW) field or an optical Raman transition \cite{Xia2015,Saffman2016}.
States $\ket{1_{\rc,\rt}}$ of the control and target atoms can be coherently
coupled to the Rydberg states $\ket{r_{\rc , \rt} }$, respectively, by focused
laser fields. The atoms excited to the Rydberg states $\ket{r_\rc}$ and
$\ket{r_\rt}$ undergo a resonant dipole-dipole exchange process
$\ket{r_\rc r_\rt} \lra \ket{a_\rc b_\rt}$ with the energy-degenerate pair
of Rydberg states $\ket{a_\rc}$ and $\ket{b_\rt}$. The dipole-dipole
interaction strength $B = C_3/x^3$ depends on the interatomic distance
$x$ and the coefficient $C_3 \propto \wp_{a_\rc r_\rc} \wp_{r_\rt b_\rt}$
is determined by the product of the dipole matrix elements $\wp \propto n^2$
of the transitions $\ket{r_\rc} \to \ket{a_\rc}$ and $\ket{r_\rt} \to \ket{b_\rt}$
between the Rydberg states with large principal quantum number $n \sim 100$,
see Appendix~\ref{ap:RSI}.

Our gate procedure is carried out in three steps (i)-(iii), similar to those 
of the Rydberg blockade protocol \cite{Jaksch2000}.
In steps (i) and (iii), resonant pulses of area 
$\theta_\mathrm{c} \equiv \int \Om_\mathrm{c} dt = \pi$ are applied 
to the control atom. For the initial state $\ket{1_\rc}$,
this amounts to the transitions $\ket{1_\rc} \to i \ket{r_\rc}$
in step (i) and $i \ket{r_\rc} \to -\ket{1_\rc}$ in step (iii).
State $\ket{0_\mathrm{c}}$ is assumed completely decoupled
from the lasers, due to transition selection rules or large transition
frequency mismatch augmented by properly shaped laser pulses.
In step (ii), a resonant pulse of area 
$\theta_\mathrm{t} \equiv \int \Om_\mathrm{t} dt = 2\pi$
is applied to the target atom resulting in the full Rabi cycle
$\ket{1_\mathrm{t}} \to i \ket{r_\rt} \to -\ket{1_\mathrm{t}}$
if the control atom is in state $\ket{0_\mathrm{c}}$. If, however,
in step (i) the control atom was excited from state $\ket{1_\mathrm{c}}$
to the Rydberg state $\ket{r_\mathrm{c}}$, the strong dipole-dipole exchange
interaction $B$ would result in the two-atom dark state
suppressing the target atom Rydberg excitation by the smooth pulse
$\Om_\mathrm{t}$, as detailed below. Here again we assume that the target
atom in state $\ket{0_\mathrm{t}}$ remains decoupled from the laser.
Steps (i)-(iii) would ideally result in a sign change ($\pi$ phase shift)
of the two-qubit states $\ket{01},\ket{10},\ket{11}$ relative to
$\ket{00}$. In combination with a Hadamard gate ($\pi/2$ rotation
on the $\ket{0_\rt} \lra \ket{1_\rt}$ transition) applied to the target qubit
before and after the phase gate, this leads to the universal \textsc{cnot}
gate between the control and target qubits \cite{MNICh2000,PLDP2007}.

\begin{figure}[t]
\includegraphics[width=8cm]{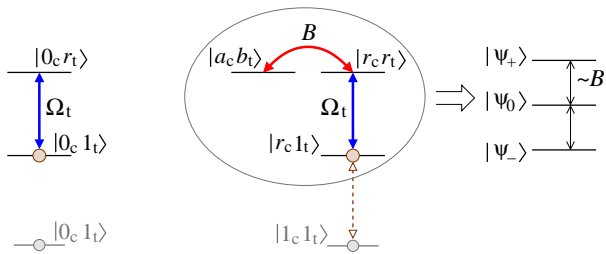}
\caption{(Color online)
Evolution of the two-atom system during step
(ii) of the Rydberg-exchange phase gate.
Left: the initial two-qubit state $\ket{0_\rc 1_\rt}$ remains unchanged
after step (i), and during step (ii) the target atom
undergoes $2\pi$ Rabi cycle via state  $\ket{0_\rc r_\rt}$.
Middle: the initial two-qubit state $\ket{1_\rc 1_\rt}$ is converted
to $\ket{r_\rc 1_\rt}$ after step (i), and during step (ii)
the resonant dipole-dipole exchange interaction
$\ket{r_\rc r_\rt} \lra \ket{a_\rc b_\rt}$ with strength $B \gg \Om_\mathrm{t}$
leads to the formation of a two-atom dark state $\ket{\psi_0}$ and
two bright states $\ket{\psi_{\pm}}$ shifted by $\pm B$ (right). 
In our protocol, state $\ket{r_\rc 1_\rt}$ adiabatically follows 
the dark state $\ket{\psi_0}$ as the target pulse $\Om_\mathrm{t}$ 
is turned on and off.}
\label{fig:alsSii}
\end{figure}

Hence, out of four possible initial two-qubit states
$\ket{00},\ket{01},\ket{10},\ket{11}$, only the last one will probe
the Rydberg-Rydberg interaction during step (ii). We shall therefore
consider in more detail the dynamics of the initial two-atom state
$\ket{1_\rc 1_\rt}$, which becomes $\ket{r_\rc 1_\rt}$ after step (i),
see Fig.~\ref{fig:alsSii}.
A pulsed laser field with Rabi frequency $\Om_\mathrm{t}$
[and area $\theta_\mathrm{t} = 2\pi$] acts
resonantly on the transition $\ket{1_\rt} \to \ket{r_\rt}$ of the target
atom, while the two-atom state $\ket{r_\rc r_\rt}$ is resonantly coupled
to state $\ket{a_\rc b_\rt}$ with strength $B$.
The Hamiltonian for the effective three-state system is then
\begin{equation}
H_{3}/\hbar = \hlf \Om_\mathrm{t} \ket{r_\rc r_\rt}\bra{r_\rc 1_\rt} +
B \ket{a_\rc b_\rt}\bra{r_\rc r_\rt} + \mathrm{H.c.} \label{eq:Ham3}
\end{equation}
This Hamiltonian has three eigenstates,
\begin{subequations}
\label{eq:eigenstates}
\begin{eqnarray}
\label{eq:eigdark}
\ket{\psi_0} &=& (B \ket{r_\rc 1_\rt} - \hlf \Om_\mathrm{t} \ket{a_\rc b_\rt})/\nu , \\
\ket{\psi_{\pm}} &=& (\hlf \Om_\mathrm{t} \ket{r_\rc 1_\rt} \mp \nu \ket{r_\rc r_\rt}
+ B \ket{a_\rc b_\rt})/\sqrt{2}\nu ,
\end{eqnarray}
\end{subequations}
with $\nu^2 \equiv B^2 + \frac{1}{4}\Om_\mathrm{t}^2$,
and the corresponding eigenvalues are $\la_0 =0$ and
$\la_{\pm} = \pm \nu$.
The zero-energy eigenstate $\ket{\psi_0}$ does not contain
the intermediate state $\ket{r_\rc r_\rt}$, and it is customary to
call it a dark state \cite{stirap}.
The energy shifted eigenstates $\ket{\psi_{\pm}}$, with
$\la_{\pm} \simeq \pm B$ ($B > \Om_t$),
are similarly called bright states.

%%%%%%%%%%%%%%%%FIGURE%%%%%%%%%%%%%%%%
%%%%%%%%%%%%%%%%%%%%%%%%%%%%%%%%

Before the laser pulse is switched on, $\Om_\mathrm{t}(0) = 0$, the
two-atom state $\ket{r_\rc 1_\rt}$ coincides with the dark state $\ket{\psi_0}$.
During the application of the pulse $\Om_\mathrm{t}(t)$,
if it is sufficiently smooth,
$|\partial_t \Om_\mathrm{t}| \ll B |\la_{\pm} - \la_0| \simeq B^2$,
the system adiabatically follows the dark state,
and the bright states $\ket{\psi_{\pm}}$ are never populated
\cite{PLDP2007,stirap}. For a smooth envelope $\Om_\mathrm{t}(t)$,
the pulse bandwidth is mainly determined by its duration
$T_\mathrm{t} \approx 2\pi /\Om_{\mathrm{t}0}$, where $\Om_{\mathrm{t}0}$
is the mean amplitude. Then the adiabatic following condition 
$|\partial_t \Om_\mathrm{t}| \approx \Om_{\mathrm{t}0} /T_\mathrm{t} \ll B^2$
reduces to $\Om_{\mathrm{t}0} \ll B$, while we assume throughout
that $\Om_{\mathrm{t}0} \gg \Ga$. 

The dark state $\ket{\psi_0(t)}$ involves instantaneous population
$P_{\mathrm{Ry}}(t) = \frac{\Om_t^2(t)}{4 B^2 + \Om_t^2(t)}$
of the two-atom Rydberg state $\ket{a_\rc b_\rt}$. During the gate time
$T_\mathrm{t}$, this population contributes 
$\Ga \int_0^{T_\mathrm{t}} P_{\mathrm{Ry}}(t) dt \approx
\pi \Ga \Om_{t0}/4 B^2$ to the decay error.
At the end of the pulse, $\Om_\mathrm{t}(t \to T_\mathrm{t}) \to 0$,
and the dark state adiabatically returns to state $\ket{r_\rc 1_\rt}$.
This state has not accumulated any phase (see below), since the adiabatically
connected eigenstate $\ket{\psi_0}$ has energy $\la_0 = 0$ for all
times $t \in [0, T_\mathrm{t}]$. Moreover, even though the double
Rydberg-excitation state $\ket{a_\rc b_\rt}$ has finite occupation
probability $P_{\mathrm{Ry}}$ while the $\Om_\mathrm{t}(t)$ pulse is on,
there is no mechanical force between the atoms, since the gradient
of energy of the two-atom eigenstate $\ket{\psi_0}$ identically
vanishes, $\partial_x \la_0 = 0 \; \forall \; t \in [0, T_\mathrm{t}]$.
Note that if the adiabatic condition is not satisfied,
after the pulse the target atom will have a residual Rydberg
population $P_{\mathrm{Ry}} \simeq \frac{\Om_{\mathrm{t}}^2}{4 B^2}$
representing a rotation error.

\section{Results and discussion}
\label{sec:results} 

The Rydberg states of atoms decay with rate $\Gamma$.
In order to minimize the error $E_\mathrm{decay} \simeq \Ga T_\mathrm{c}$
due to decay of the control atom, we thus need to accomplish
steps (i) and (iii) in shortest possible times using
strong pulses of mean amplitude $\Om_\mathrm{c0}$ and duration
$T_\mathrm{c} = \pi / \Om_\mathrm{c0}$. Similarly, to minimize the error
$E_\mathrm{decay} \simeq \Ga T_\mathrm{t}$ due to the Rydberg state
decay of either the control atom or the target atom during step (ii),
$T_\mathrm{t} \approx 2\pi / \Om_\mathrm{t0}$, we should take the mean
Rabi frequency $\Om_\mathrm{t0}$ as large as possible, but it should
still be smaller than the interaction $B$, to satisfy the
adiabatic following condition detailed above.
It then follows that, for the gate performed with a smooth (adiabatic)
pulse $\Om_{\mathrm{t}}$, during step (ii) the intrinsic error probability
averaged over all the possible two-qubit inputs is
\begin{equation}
E \simeq \frac{\pi \Ga}{4} \left[\frac{5}{\Om_{\mathrm{t0}}}
+ \frac{\Om_{\mathrm{t0}}}{4 B^2}\right] .
\end{equation}
If we minimize $E$ with respect to $\Om_{\mathrm{t0}}$,
we find $E=\frac{\sqrt{5} \pi \Ga}{4 B}$
for $\Om_{\mathrm{t0}} = 2\sqrt{5} B$, which, however,
violates the adiabatic criterion. Furthermore,
such a large Rabi frequency $\Om_{\mathrm{t0}} \gtrsim B$ is difficult
to achieve experimentally for high-lying Rydberg states. 
Instead, we can choose $\Om_{\mathrm{t0}} = \alpha B$ with $\alpha \ll 1$, 
obtaining $E \simeq \frac{5 \pi \Ga}{4 \Om_{\mathrm{t0}}}$, or
$E \simeq \eta \frac{\Ga}{B}$ with $\eta \simeq \frac{5 \pi}{4 \alpha}$.

We can estimate the minimum attainable error as follows.
In a cryogenic environment with no black-body radiation, 
the radiative lifetime of the $ns,np,\ldots$
Rydberg states of the alkali atoms is given by
$\tau =1/\Ga \approx 10^{-9} n^3\:$sec \cite{Gounand1979,RydAtoms,Beterov2009}.
The strongest interaction is achieved with the dipole-dipole
potential $B \simeq \frac{1}{\hbar} \frac{\wp^2}{4 \pi \epsilon_0 x^3}$,
where $\wp \sim a_0 e n^2$ is the dipole moment of the atom in
the Rydberg state. At the interatomic distance of $x = 3-5\:\mu$m,
we then have $B \simeq 100 n^4\:$rad/s. To avoid population leakage to
other Rydberg states, this interaction strength should be smaller than 
the level separation between neighboring $n$ states, 
$\de \omega_{\mathrm{F}} \sim 2 \, \mathrm{Ry} \, n^{-3} > B$ 
(see Appendix~\ref{ap:RSI}).
This then leads to the condition $n \lesssim 100$, which also follows
from the requirement that the Rydberg electron clouds (of size $\sim a_0 n^2$)
of neighboring atoms do not overlap.
We thus obtain $B \tau \approx 10^{-7} n^7 \sim 10^7$ for $n \sim 100$.
Choosing $\alpha = \Om_{\mathrm{t}}/B \simeq 0.1$ ($\eta \simeq 40$), the 
minimal error probability is $E \simeq \eta \frac{\Ga}{B} \lesssim 10^{-5}$.

We note that general arguments \cite{Wesenberg2007} put a lower
limit $E \simeq \frac{2\Ga}{B}$ on the gate error due to
decay of the interacting excited states, which is
an order of magnitude smaller than in our case. This is
due to our requirement of adiabatic, i.e., slow, evolution
of the system to avoid population leakage to undesired
states. To speed up the gate and reduce the accumulated decay
probability of the Rydberg states, one may resort to recently
developed ``short-cut to adiabaticity'' schemes \cite{Muga2013}.
In particular, using the so-called derivative removal by
adiabatic gate (DRAG) pulses \cite{DRAG,Theis2016}, may
accelerate the gate by operating in the regime of
$\Om_{\mathrm{t}0} \sim B$, provided the necessary laser intensities
can be achieved.

As we discuss in Appendices~\ref{ap:RSI} and \ref{ap:Ham}, the Rydberg
product states $\ket{r_\rc r_\rt}$ and $\ket{a_\rc b_\rt}$ might experience
non-resonant dipole-dipole couplings to other Rydberg  pair states.
While population leakage to these states is reduced in the adiabatic
regime, dispersive coupling with these states will result in second
order (van der Waals) energy shifts $\beta_{rr}$ and $\beta_{ab}$
of states $\ket{r_\rc r_\rt}$ and $\ket{a_\rc b_\rt}$.
The two-atom dark state $\ket{\psi_0}$ does not involve
the population of the intermediate state $\ket{r_\rc r_\rt}$ and
is insensitive to its energy shifts $\beta_{rr}$. But the energy shift
$\beta_{ab}$ of state $\ket{a_\rc b_\rt}$ perturbs the dark state $\ket{\psi_0}$.
This perturbation will not result in the coupling of $\ket{\psi_0}$
to the bright states $\ket{\psi_{\pm}}$ as long as the shift $\beta_{ab}$
is small compared to the exchange interaction strength $B$, since the latter
determines the energy splitting of the bright eigenstates and thereby
the width of the dark resonance. Yet, during the gate execution the small
but finite population of the energy-shifted state $\ket{a_\rc b_\rt}$
will result in a phase shift of the dark state,
$\phi = \int_0^{T_{\mathrm{t}}} \beta_{ab} P_{\mathrm{Ry}}(t) dt$.
This phase can be amended, as described in \cite{Zhang2012} and in
Appendix~\ref{ap:RBG}. Otherwise, we can tune the F\"orster frequency
defect $\delta \omega$ for the transition
$\ket{r_\rc r_\rt} \to \ket{a_\rc b_\rt}$ to exactly compensate this
level shift, $\delta \omega = - \beta_{ab}$, as discussed
in Appendix~\ref{ap:Ham}. Then the phase shift $\phi$ will vanish
for any resonant pulse $\Om_{\mathrm{t}}(t)$.

%%%%%%%%%%%%%%%%FIGURE%%%%%%%%%%%%%%%%
\begin{figure}[t]
\includegraphics[width=8.7cm]{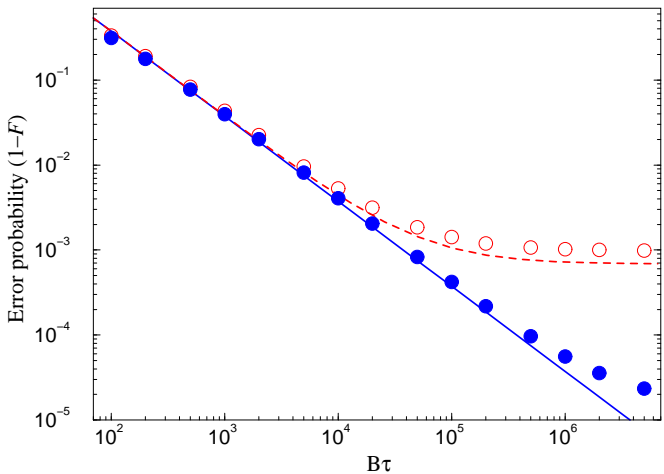}
\caption{(Color online)
Error probabilities $E=(1-F)$ of the phase gate,
averaged over all the input states, versus $B \tau$.
Filled blue circles correspond to smooth $2\pi$ laser pulses
$\Om_\mathrm{t}(t)$ applied to the target qubit, while empty red circles
to non-adiabatic (square) pulses of the same area and duration. The results
are obtained by numerical simulation of the dynamics of the two-atom
system with a non-Hermitian Hamiltonian, as described in Appendix~\ref{ap:Ham}.
The solid blue line shows $E = \eta/(B \tau)$ and the dashed red line
shows  $E = \eta/(B \tau) + \alpha^2/16$.
All Rydberg states decay with the same rate $\Ga$, while
qubit states $\ket{0}$ and $\ket{1}$ do not decay.
We used $\alpha \equiv \Om_\mathrm{t0}/B = 0.10472$ ($\eta = 37.5$)
resulting in the non-adiabatic transition errors
$E_{\mathrm{na}} \simeq 2 \times 10^{-6}$ for smooth (shifted Gaussian) pulses.}
\label{fig:errCZ}
\end{figure}
%%%%%%%%%%%%%%%%%%%%%%%%%%%%%%%%

We have verified the above qualitative results by exact
numerical simulations of the dynamics of the two-atom system,
as detailed in Appendix~\ref{ap:Ham}. We use smooth $2\pi$ laser
pulses $\Om_\mathrm{t}(t)$ applied to the target qubit during step (ii)
and stronger $\pi$-pulses $\Om_\mathrm{c0} = 4 \Om_\mathrm{t0}$ applied
to the control atom in steps (i) and (iii).
In Fig.~\ref{fig:errCZ} we show the error probabilities 
$E =1-F$ of the phase gate. The gate fidelity $F$ is obtained
by averaging over all possible two-qubit input states,
as described in Appendix~\ref{ap:Ham} and Ref. \cite{Pedersen2007}. 
We observe that the error follows approximately the linear
scaling $E=\eta/(B\tau)$, but for large values of $B \tau \gtrsim 10^6$,
the numerically obtained error probabilities start to deviate from 
the analytic estimate of the error during only step (ii). 
This is due the contribution of the additional error $\sim 10^{-5}$ stemming 
from the decay of the control qubit $E_\mathrm{decay} \simeq \pi \Ga/\Om_\mathrm{c0}$ 
during steps (i) and (iii), non-adiabatic transitions and leakage from 
the dark state $E_{\mathrm{na}}$, and imperfect phase compensation.
Nevertheless, we obtain that the average fidelity reaches 
$F = 0.99995$ for $B \tau \gtrsim 10^6$.

For comparison, we also show in Fig.~\ref{fig:errCZ} the results of
simulations for the gate performed with square $2\pi$-pulses acting
on the target qubit. Now, starting from the values of $B\tau \geq 10^5$,
the error probability significantly deviates from the linear scaling,
which is due to the breakdown of adiabaticity leading to the residual
Rydberg population of the target atom $\sim \Om_\mathrm{t}^2/16 B^2$.
There is also sizable population leakage to other Rydberg
states not accounted for by the analytic estimate of the gate error.
We note that this error is of the same magnitude as the rotation
error due to imperfect Rydberg blockade \cite{Zhang2012}.
As we show in Appendix~\ref{ap:RBG}, this error can also be avoided
in the usual Rydberg blockade scenario \cite{Jaksch2000}, using
either adiabatic pulses to excite and de-excite the target atom,
or by applying a square pulse of proper amplitude
$\Om_{\mathrm{t}} = \frac{B_{\mathrm{sh}}}{\sqrt{4k^2 - 1}}$ ($k \in \mathbf{N}$)
which accomplishes both a full resonant $2\pi$ Rabi cycle and
a full precession of the two-level Bloch vector with the
generalized off-resonant Rabi frequency
$\bar{\Om} \equiv \sqrt{B_{\mathrm{sh}}^2 + \Om_{\mathrm{t}}^2}$.
Such pulses, however, contribute interaction phases
to the quantum state amplitudes which can only be compensated
for if we know precisely the interatomic distance and thereby
the interaction strength $B_{\mathrm{sh}}$. Moreover, during the standard
blockade or interaction gates with dispersive interatomic interaction
(static dipole-dipole or van der Waals level-shift $B_{\mathrm{sh}}$),
for any non-vanishing probability of double  Rydberg excitation,
the atoms are subject to forces due to the spatially dependent
potential.

\section{Conclusions}

In summary, we have examined the phase gate performance
using strong resonant dipole-dipole interactions between pairs
of atoms in Rydberg states. 
Our gate assumes atomic level and laser excitation schemes 
which are similar to the ones used in current experiments.
Employing adiabatic excitation of
the Rydberg states of atoms with smooth laser pulses,
we find favorable scaling of the intrinsic gate errors
$E \propto (\Ga/B)$ with the ratio of the Rydberg state
decay rate $\Ga$ to the interaction strength $B$, which
should be contrasted with the optimized error probability
$E \propto (\Ga/B)^{2/3}$ obtained in the previous studies
\cite{Saffman2005,LIMSKM2011,Zhang2012} with non-adiabatic pulses.
The better scaling of the gate error probability is due to nearly 
complete elimination of the residual Rydberg excitation of the 
imperfectly blockaded atom. 
The corresponding gate fidelity can reach $F > 0.9999$ for $B/\Ga \simeq 10^6$.
The ultimate limit on gate fidelity depends on the value of $B/\Ga$ 
and the ability to suppress other technical sources of errors. 
While the analysis of Sec.~\ref{sec:results} shows that generically 
$B \tau \lesssim 10^7$, the precise limit may be higher. 
Thus, for cases 1-5 in Appendix~\ref{ap:RSI}, we find that $B \tau$ 
can be as high as $4 \times 10^7$ in a cryogenic environment at 4K,
which implies that a fidelity of $F = 1- 10^{-5}$ is feasible. 

We have focused in this paper on the intrinsic gate error $E$ due to 
the decay of the Rydberg states and their finite interaction strength. 
In any real experiment, however, there will also be technical errors, 
due to, e.g., the laser phase fluctuations and Doppler shifts leading 
to dephasing $\gamma$ of the atomic transition, and variations of
the laser pulse duration or amplitude leading to pulse area uncertainty 
$\delta \theta$. If we require that $\gamma \lesssim \Gamma/2$ and 
$\delta \theta \lesssim \sqrt{E}$, these errors will not exceed 
the intrinsic error $E$ and adversely affect the system.

\begin{acknowledgments}
This work was supported by the US ARL-CDQI program
through cooperative agreement W911NF-15-2- 0061,
the EU H2020 FET-Proactive project RySQ, and the Villum Foundation.
\end{acknowledgments}

\appendix

\section{Resonant dipole-dipole interactions of Rydberg-state atoms}
\label{ap:RSI}

In the main text, we discuss the realization of a quantum
gate using a two-atom dark-state resonance which employs a resonant
dipole-dipole exchange interaction between a pair of Rydberg atoms in a state
$\ket{r_\rc r_\rt}$  and a state $\ket{a_\rc b_\rt}$ with
the same energy, $\veps_{r_\rc} + \veps_{r_\rt} = \veps_{a_\rc} + \veps_{b_\rt}$
(see Fig.~\ref{fig:als}).
The most obvious choice of Rydberg states that exhibit strong
resonant exchange interaction is $\ket{r_\rc} = \ket{b_\rt} \equiv ns_{1/2}$
and $\ket{r_\rt}= \ket{a_\rc} \equiv np_{3/2}$ with a large principal quantum
number $n \sim 100$. State $\ket{1_\rc}$ of the control atom can then
be coupled to the Rydberg state $\ket{r_\rc}$ by a two-photon
transition via a virtual intermediate state involving two optical
(or a UV and a MW) photons. State $\ket{1_\rt}$ of the target atom can
be coupled to the Rydberg state $\ket{r_\rt}$ by a single UV photon.

While being automatically resonant
for the $\ket{r_\rc r_\rt} \lra \ket{a_\rc b_\rt}$ transition, this choice of Rydberg states, 
however, presents problems associated with near-resonant coupling to other, unwanted states.
Recall that, in the absence of external electric or magnetic fields,
the energies of the Rydberg states are given by
$\veps_{nl} \equiv \hbar \omega_{nl} = -\frac{\mathrm{Ry}}{(n-\delta_l)^2}$,
where $\mathrm{Ry}$ is the Rydberg constant and $\delta_l$
is the quantum defect for the angular momentum states with 
$l=s,p,\ldots$ \cite{rydQIrev,RydAtoms}. 
For large $n$, we then obtain that the frequency mismatch $\delta \omega_{rr(ab)}
= \omega_{ns} + \omega_{np} - \omega_{(n+dn)p} - \omega_{(n-dn)s}$ 
for transitions from $\ket{ns, np}$ to unwanted states
$\ket{(n+dn)p,(n-dn)s}$ scales as
$\delta \omega_{rr} \simeq \delta \omega_{\mathrm{F}}
\frac{3(dn +\delta_S - \delta_P)}{n} \ll \delta \omega_{\mathrm{F}}$,
rather than the familiar F\"orster defect
$\delta \omega_{\mathrm{F}} = \mathrm{Ry}\frac{2dn}{n^3}$.

\subsection{Stark-tuned F\"orster resonances}

%%%%%%%%%%%%%%%%FIGURE%%%%%%%%%%%%%%%%
\begin{figure}[t]
\includegraphics[width=5.3cm]{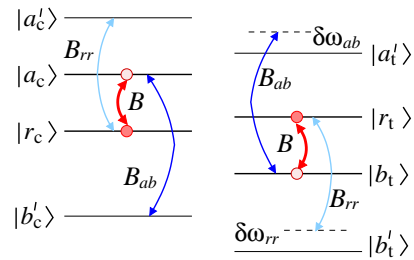}
\caption{(Color online)
Energy level structure of Rydberg states for Stark tuned
F\"orster interaction.
Resonant lasers couple the qubit states $\ket{1_{\rc,\rt}}$ of the control
and target atoms to the corresponding Rydberg states $\ket{r_{\rc,\rt}}$
(see Fig.~\ref{fig:als}).
A static electric field shifts the Rydberg states making
the two-atom states $\ket{r_\rc r_\rt}$ and $\ket{a_\rc b_\rt}$
energy degenerate and resonantly coupled via exchange interaction
with strength $B$.
The forward leakage channel couples $\ket{r_\rc r_\rt}$ to $\ket{a_\rc' b_\rt'}$
with strength $B_{rr}$ and energy defect $\delta \omega_{rr}$.
The backward leakage channel couples $\ket{a_\rc b_\rt}$ to
$\ket{b_\rc' a_\rt'}$ with strength $B_{ab}$ and energy defect
$\delta \omega_{ab}$.}
\label{fig:StarkForster}
\end{figure}
%%%%%%%%%%%%%%%%%%%%%%%%%%%%%%%%

We can mitigate this problem by resorting instead to the Stark 
tuned F\"orster resonances, which have been demonstrated in 
Refs. \cite{Ryabtsev2010,Ravets2014}. In principle, Stark tuning with an 
appropriate external static electric field $E_{\mathrm{St}}$ can render 
any pair of two-atom states $\ket{r_\rc r_\rt}$ and $\ket{a_\rc b_\rt}$ 
degenerate. These two-atom states may still couple to many other states 
resulting in leakage and gate errors, and our task is to search for state 
combinations minimizing this leakage. We consider two types of leakage 
channels, as shown in Fig.~\ref{fig:StarkForster}:
forward leakage where $\ket{r_\rc r_\rt}$ couples to states
$\ket{a_\rc' b_\rt'}$ with strength $B_{rr}$ and F\"orster defect
$\delta \omega_{rr}$; and backward leakage that couples
$\ket{a_\rc b_\rt}$ to states $\ket{b_\rc' a_\rt'}$ with strength $B_{ab}$
and F\"orster defect $\delta \omega_{ab}$. We neglect higher order leakage
processes where the states $\ket{a_\rc' b_\rt'}$ and $\ket{b_\rc' a_\rt'}$
couple to $\ket{b_\rc'' a_\rt''}$ and $\ket{a_\rc'' b_\rt''}$ etc.

We have searched for suitable Cs atom pair Rydberg states by choosing
$\ket{r_\rc r_\rt}$ and $\ket{a_\rc b_\rt}$ and then finding the static
Stark field $E_{\mathrm{St}}$, directed along the quantization axis,
that makes the pairs energy degenerate. We used standard expressions
for the scalar and tensor polarizabilities \cite{LeKien2013}
in the fine structure basis calculated by summing over all dipole allowed
transitions over a range of principal quantum numbers of $\pm 20$ from
each state. Small hyperfine corrections have been neglected.
Radial matrix elements between Rydberg states were calculated using
the quasi-classical (WKB) approximation \cite{Kaulakys1995} with quantum
defect values taken from \cite{Lorenzen1984,Weber1987}. The effective
matrix elements were not corrected for state mixing due to $E_{\mathrm{St}}$.

We checked all possible dipole allowed transitions from $\ket{r_\rc r_\rt}$
and $\ket{a_\rc b_\rt}$ to the leakage states with the change of principal
quantum number up to $\pm 5$ from the resonant states. The F\"orster energy
defects of leakage state pairs were calculated for $E_{\rm St}$ corresponding
to the $\ket{r_\rc r_\rt} \lra \ket{a_\rc b_\rt}$ resonance. The two atom
dipolar coupling coefficient is given by the general expression
\begin{eqnarray}
C_3 &=& - \frac{e^2}{4\pi\epsilon_0}
\frac{\sqrt6 (4\pi)^{3/2}}{3\sqrt5 } \sum_{M,q} (-1)^M C_{1,q,1,-M-q}^{2,-M}
\nonumber \\
& & \times Y_{2,M}(\hat{\varepsilon})(rY_{1,q})^{(\rc)}(rY_{1,-M-q})^{(\rt)} ,
\end{eqnarray}
where $C_{....}^{..}$ is a Clebsch-Gordan coefficient, $Y_{2,M}$ is a spherical
harmonic, $\hat{\varepsilon}$ is a unit vector pointing from atom $(\rc)$ to
atom $(\rt)$, and $(rY_{1,q})^{(\rc)}$ and $(rY_{1,-M-q})^{(\rt)}$ are the relative
electron positions for each atom in spherical coordinates. We evaluate
the above expression for $\hat{\varepsilon}$ perpendicular to 
the quantization axis. This corresponds to the geometry
of a planar array of atoms that may be individually addressed while having
isotropic interactions in the plane, as in Ref.~ \cite{Maller2015}.
In this geometry, the selection rules for the dipole-dipole interaction
are $\Delta M = 0, \pm 2$. Fewer leakage channels occur with
$\hat{\varepsilon}$ along the quantization axis which limits the interactions
to $\Delta M=0$, but such a geometry is less convenient for a multi-qubit
implementation of quantum information processing with trapped neutral atoms.

\begin{widetext}

%%%%%%%%%%%%%%%%%TABLE%%%%%%%%%%%%%%%%%
\begin{table*}[h]
\caption{Resonant Rydberg atom pair states $\ket{r_\rc r_\rt}$ and $\ket{a_\rc b_\rt}$,
specified as $nl_jm$, with the corresponding dipole-dipole interaction coefficient $C_3$,
parameter $B\tau$ at interatomic separation $x= 3\:\mu$m and temperatures 300K and 4K, 
and the strength of the tuning Stark field $E_{\mathrm{St}}$.}
\begin{ruledtabular}
\begin{tabular}{cccccccc}
\textrm{case} & $\ket{r_\rc}$ & $\ket{r_\rt}$ & $\ket{a_\rc}$ & $\ket{b_\rt}$ 
& $C_3\:$(\textrm{GHz}$\:\mu$\textrm{m}$^3$) & $B\tau (\times 10^6)$  &  $E_{\mathrm{St}}\:$(\textrm{V/m})\\
  &                &               &                &                &       &    300K,  4K    & \\
\colrule
1 & $109s_{1/2}1/2$ & $101s_{1/2}1/2$ & $109p_{3/2}3/2$ & $101p_{3/2}3/2$ & -64.4 & 6.5, 32.6 & 15.4 \\
2 & $112s_{1/2}1/2$ & $101s_{1/2}1/2$ & $111p_{3/2}3/2$ & $101p_{3/2}3/2$ &  65.3 & 6.8, 35.5 & 5.36 \\
3 & $105s_{1/2}1/2$ &  $94s_{1/2}1/2$ & $105p_{3/2}3/2$ &  $94p_{3/2}3/2$ & -51.4 & 4.7, 23.1 & 20.1 \\
4 & $112p_{3/2}3/2$ & $101p_{3/2}3/2$ & $112s_{1/2}1/2$ & $101s_{1/2}1/2$ & -68.2 & 7.1, 38.5 & 14.2 \\
5 &  $95p_{3/2}3/2$ &  $84p_{3/2}3/2$ &  $95s_{1/2}1/2$ &  $84s_{1/2}1/2$ & -33.0 & 2.7, 10.7 & 34.7 \\
\end{tabular}
\end{ruledtabular}
\label{tab:StarkForster1}
\end{table*}
%%%%%%%%%%%%%%%%%%%%%%%%%%%%%%%%%%%%

%%%%%%%%%%%%%%%%%TABLE%%%%%%%%%%%%%%%%%
\begin{table*}[h]
\caption{The dominant forward and backward leakage states, corresponding to the largest
absolute values of parameters $\beta_{rr} = B_{rr}^2/\delta \omega_{rr}$ and
$\beta_{aa} = B_{aa}^2/\delta \omega_{aa}$, for the cases 1-5 in Table \ref{tab:StarkForster1}.
The last column gives the population missing from state $\ket{r_{\rc} 1_\rt}$ after step (ii)
as found from numerical integration of the Schr\"odinger equation with Hamiltonian (\ref{eq:Ham5})
and the laser parameters given there.}
\begin{ruledtabular}
\begin{tabular}{cccccccccc}
\textrm{case} & $\ket{a_\rc'}$  & $\ket{b_\rt'}$ & $B_{rr}/B$ & $\delta \omega_{rr}/2\pi \:$(MHz)
              & $\ket{b_\rc'}$  & $\ket{a_\rt'}$ & $B_{ab}/B$ & $\delta \omega_{ab}/2\pi\:$(MHz)
              & $1-P_{\ket{r1}}(T_t)$\\
\colrule
1 & $109p_{1/2}(-1/2)$ & $101p_{3/2}(-1/2)$ & -0.49 &  65 & $108d_{5/2}5/2$  &  $99d_{5/2}5/2$ & -0.64 &  190 & $1.4\times 10^{-5}$ \\
2 & $111p_{1/2}(-1/2)$ & $101p_{1/2}(-1/2)$ &  0.66 &-259 & $110d_{5/2}5/2$  & $100d_{5/2}5/2$ & -2.17 & 1990 & $1.7\times 10^{-6}$ \\
3 & $105p_{1/2}(-1/2)$ &  $94p_{3/2}(-1/2)$ & -0.49 & 247 & $104d_{5/2}5/2$  &  $92d_{5/2}5/2$ & -0.64 & -185 & $5.5\times 10^{-5}$ \\
4 & $111d_{5/2}(5/2)$  &  $99d_{5/2}(5/2)$  & -0.64 & -75 & $112p_{1/2}(-1/2)$& $101p_{3/2}(-1/2)$ & -0.49 & 171 & $9.5 \times 10^{-6}$\\
5 &  $94d_{5/2}(5/2)$  &  $82d_{5/2}(5/2)$  & -0.64 &-478 &  $95p_{3/2}3/2$  & $84p_{1/2}(-1/2)$ & 0.28 & 122 &  $6.4 \times 10^{-6}$\\
\end{tabular}
\end{ruledtabular}
\label{tab:StarkForster2}
\end{table*}
 %%%%%%%%%%%%%%%%%%%%%%%%%%%%%%%%%%%%
\end{widetext}

In Tables \ref{tab:StarkForster1}, and \ref{tab:StarkForster2}
we show several possible choices of the Rydberg atom pair
states $\ket{r_\rc r_\rt}$ and $\ket{a_\rc b_\rt}$. One possibility (cases 1,2,3)
is $\ket{r_\rc r_\rt} = \ket{n_{\rc} s_{1/2},m=1/2; n_{\rt}s_{1/2},m=1/2}$,
with the Stark field set for resonance with
$\ket{a_\rc b_\rt} = \ket{n_{\rc} p_{3/2},m=3/2; n_{\rt} p_{3/2},m=3/2}$.
The laser excitation of the Rydberg states $\ket{r_\rc}$ and $\ket{r_\rt}$
from the Cs ground state requires two-photon transitions.
The forward leakage channels from $\ket{r_\rc r_\rt}$ are to $p_{1/2}$
or $p_{3/2}$ states. The backward leakage channels couple $\ket{a_\rc b_\rt}$
to either two $s$ states with different $n$, two $d$ states,
or an $s$ and a $d$ state.
Another possibility (cases 4,5) is
$\ket{r_\rc r_\rt} =\ket{n_{\rc} p_{3/2}, m=3/2; n_{\rt} p_{3/2}, m=3/2}$
tuned to resonance with
$\ket{a_\rc b_\rt} = \ket{n_{\rc} s_{1/2},m=1/2; n_{\rt}s_{1/2},m=1/2}$.
The Rydberg states  $\ket{r_\rc}$ and $\ket{r_\rt}$ can now be reached with
one UV photon starting from the Cs ground state. Although the fine structure
splitting between $np_{3/2}$ and $np_{1/2}$ states is small at large $n$,
undesired coupling to $np_{1/2}$ is strongly suppressed in the heavy alkali
atoms \cite{Fermi1930}. Note that state $\ket{a_\rc b_\rt}$ can couple strongly
with $\ket{n_{\rc} p_{1/2}; n_{\rc}p_{1/2}}$, but the energy separation of the $np$
fine structure states is increased in the presence of a Stark field.

\subsection{Numerical estimates of the leakage errors}

To estimate the gate error due to the leakage of population of states
$\ket{r_\rc r_\rt}$ and $\ket{a_\rc b_\rt}$ to the non-resonant states
$\ket{a_\rc' b_\rt'}$ and $\ket{b_\rc' a_\rt'}$, we solve the Schr\"odinger
equation for the two-atom system subject to the Hamiltonian
\begin{eqnarray}
H_{5}/\hbar &=& H_{3}/\hbar + \delta \omega_{rr} \ket{b_\rt'}\bra{b_\rt'}
+ \delta \omega_{ab} \ket{a_\rt'}\bra{a_\rt'}
\nonumber \\ & &
+ B_{rr} \ket{a_\rc' b_\rt'}\bra{r_\rc r_\rt}
+ B_{ab} \ket{b_\rc' a_\rt'}\bra{a_\rc b_\rt} + \mathrm{H.c.}
\qquad \label{eq:Ham5}
\end{eqnarray}
where $H_{3}$ is the Hamiltonian of Eq.~(\ref{eq:Ham3}).
Starting with the initial state $\ket{r_\rc 1_\rt}$, we apply a smooth
pulse of duration $T_\mathrm{t}=2\pi/\alpha B$ and
a (shifted) Gaussian temporal shape
$\Om_\mathrm{t}(t) = A [e^{-(t-T_\mathrm{t}/2)^2/(2\sigma^2)}-e^{-(T_\mathrm{t}/2)^2/(2\sigma^2)}]$,
where $A$ is chosen such that 
$\theta_\mathrm{t} = \int_0^{T_\mathrm{t}} \Om_\mathrm{t} dt = 2\pi$, and
we take $\sigma = T_\mathrm{t}/5$ leading to the peak Rabi frequency
$\Omega_{t}(T_\mathrm{t}/2)=2.1 \times 2\pi/T_\mathrm{t}$.
Numerical simulations were performed with $B/2\pi=350\:$MHz, $\alpha=0.1$,
$T_t=29\:$ns, and a peak Rabi frequency of $\Omega_{t}(T_\mathrm{t}/2)/2\pi=74\:$MHz.
In Table \ref{tab:StarkForster2}, last column, we show the population missing
from state $\ket{r_{\rc} 1_\rt}$ at the end of the pulse. In all cases, the population
rotation error is in the range of $2 \times 10^{-6} - 5\times 10^{-5}$, with the
smallest error obtained for case 2. A full analysis including all
the leakage channels will undoubtedly show larger errors. The results presented 
here, however, account for the dominant leakage and we are optimistic that 
nonadiabatic effects may further be reduced by pulse shaping \cite{Theis2016}.

\section{Details of the numerical calculations}
\label{ap:Ham}

In Fig. \ref{fig:errCZ} we present the results of numerical simulations
of the complete phase gate between the control and target qubits represented
by the atoms. The system we simulate consists of two six-level atoms described 
by the Hamiltonian of the form
\begin{equation}
H = H_{\mathrm{MW}} + H_{\mathrm{L}} + H_{\mathrm{Ry}}. \label{eq:HamFull}
\end{equation}
Here the first term describes the qubit states of the atoms, $\ket{0_{\rc,\rt}}$
and$ \ket{1_{\rc,\rt}}$, and their manipulation by the microwave fields,
\begin{eqnarray}
H_{\mathrm{MW}}/\hbar &=& - \Delta^{(\rc)}_{\mathrm{MW}}(t) \ket{0_\rc}\bra{0_\rc} - \Delta^{(\rt)}_{\mathrm{MW}}(t) \ket{0_\rt}\bra{0_\rt}
\nonumber \\ & &
\hlf \Omega_{\mathrm{MW}}(t) (\ket{1_\rc}\bra{0_\rc} + \ket{1_\rt}\bra{0_\rt}) + \mathrm{H.c.} \qquad
\end{eqnarray}
where $\Omega_{\mathrm{MW}}(t)$ is the Rabi frequency of the pulsed microwave field seen by both atoms,
and the selectivity is provided by setting the detuning $\Delta^{(\rc,\rt)}_{\mathrm{MW}}(t)$ of each
atom to either $\Delta_{\mathrm{MW}} =0$ or $\Delta_{\mathrm{MW}} \gg |\Omega_{\mathrm{MW}}|$. This can
be done by using non-resonant laser light tightly focused onto the selected atom to induce an
ac Stark shift of one of the qubit states \cite{Xia2015}. We always start with the two atom state
$\ket{0 0}$ and prepare one of the four input states $\ket{00}$, $\ket{01},\ket{10},\ket{11}$
by applying a microwave $\pi$ pulse, with the atom(s) required to switch to state $\ket{1}$
being resonant ($\Delta_{\mathrm{MW}} =0$), and the atom(s) required to remain in $\ket{0}$
being strongly detuned ($\Delta_{\mathrm{MW}} =100 |\Omega_{\mathrm{MW}}|$). The Hadamard gates on the atoms can also be performed in the same way, with the $\pi/2$ microwave pulse having the
phase $\arg(\Omega_{\mathrm{MW}}) = -\pi/2$.

The second term of Eq.~(\ref{eq:HamFull}) describes the resonant laser coupling of the qubit
states $\ket{1_{\rc,\rt}}$ of the control and target atoms to the Rydberg states
$\ket{r_{\rc,\rt}}$,
\begin{equation}
H_{\mathrm{L}}/\hbar = \hlf \Omega_{\rc}(t) \ket{r_{\rc}}\bra{1_{\rc}} + \hlf \Omega_{\rt}(t) \ket{r_{\rt}}\bra{1_{\rt}} + \mathrm{H.c.}
\end{equation}
The lasers are focused onto the atoms, and we apply strong $\pi$ pulses
to the control atom in steps (i) and (iii) of the protocol, and a smooth $2\pi$
pulse to the target atom in step (ii). The temporal shape of the target laser 
pulse is 
\begin{equation}
\Om_\mathrm{t}(t) = A [e^{-(t-T_\mathrm{t}/2)^2/(2\sigma^2)}-e^{-(T_\mathrm{t}/2)^2/(2\sigma^2)}] \label{eq:Gausspls}
\end{equation}
with
\[
A = \frac{e^{T_\rt^2/(8\sigma^2)} -1}{1-e^{-(T_\rt/2)^2/(2\sigma^2)} }
\frac{\sqrt{2\pi} }{\sigma e^{T_\rt^2/(8\sigma^2)}\mathrm{erf}(T_\rt/(2^{3/2}\sigma)) -T_\rt }.
\]
The pulse duration is $T_\mathrm{t}=2\pi/\Om_\mathrm{t0}$ with 
$\Om_\mathrm{t0}/B =\alpha =0.10472$, and we take $\sigma = T_\mathrm{t}/5$. 
The shifted Gaussian pulses of the form (\ref{eq:Gausspls}) have 
the advantage of being smooth, finite-duration pulses that can be readily 
implemented experimentally. The time intervals between the laser pulses 
in steps (i), (ii), (iii) are set to $T_\mathrm{t}/20$. 

Finally, the last term of Eq.~(\ref{eq:HamFull}) describes the Rydberg states 
of atoms and their interactions,
\begin{eqnarray}
H_{\mathrm{Ry}}/\hbar &=&  \delta \omega \ket{b_\rt}\bra{b_\rt} + \delta \omega_{rr} \ket{b_\rt'}\bra{b_\rt'}
+ \delta \omega_{ab} \ket{a_\rt'}\bra{a_\rt'}
\nonumber \\ & &
+ ( B \ket{a_\rc b_\rt}\bra{r_\rc r_\rt} + B_{rr} \ket{a_\rc' b_\rt'}\bra{r_\rc r_\rt}
\nonumber \\ & & \qquad
+ B_{ab} \ket{b_\rc' a_\rt'}\bra{a_\rc b_\rt} + \mathrm{H.c.} ),
\qquad \label{eq:HamRyd}
\end{eqnarray}
where we  include the F\"orster defect $\delta \omega$ on
the Stark-tuned transition $\ket{r_\rc r_\rt} \lra \ket{a_\rc b_\rt}$.
When the transitions to the unwanted states $\ket{a_\rc' b_\rt'}$
and $\ket{b_\rc' a_\rt'}$ are non-resonant, i.e., the corresponding
F\"orster defects are large, $\delta \omega_{rr(ab)} > B_{rr(ab)}$,
the leakage from the two-atom dark state is suppressed.
Yet, the non-resonant couplings induce
second-order level shifts of states $\ket{r_\rc r_\rt}$ and
$\ket{a_\rc b_\rt}$, given by $\beta_{rr} = B_{rr}^2/\delta \omega_{rr}$
and $\beta_{ab} = B_{ab}^2/\delta \omega_{rr}$, respectively.
The dark state $\ket{\psi_0}$ is insensitive to the small energy
shift $\beta_{rr}$ of the intermediate state $\ket{r_\rc r_\rt}$.
But the energy shift $\beta_{ab}$ of state $\ket{a_\rc b_\rt}$ perturbs
the dark resonance. If $\beta_{ab}$ is small compared to the energy
splitting $\pm B$ of the bright eigenstates $\ket{\psi_\pm}$
(see Sec.~\ref{sec:tads}), they will not be populated from $\ket{\psi_0}$
under the adiabatic condition. However, during the phase gate sequence
starting from state $\ket{r_\rc 1_\rt}$, the small but finite population
$P_{\mathrm{Ry}}(t) = \frac{\Om_t^2(t)}{4 B^2 + \Om_t^2(t)}$ of state
$\ket{a_\rc b_\rt}$ will result in the dark state accumulating the phase
$\phi = \int_0^{T_{\mathrm{t}}} \beta_{ab} P_{\mathrm{Ry}}(t) dt \approx
\kappa \frac{\pi^2 \beta_{ab}}{B^2T_{\mathrm{t}}}$, where $\kappa = O(1)$
depends on the $\Om_{\mathrm{t}}(t)$ pulse shape. For example,
$\kappa = \frac{\pi^2}{8} \simeq 1.23$ for the pulse
$\Om_\mathrm{t}(t) = \hlf \pi \Om_\mathrm{t0} \sin(\pi t/T_\mathrm{t})$,
and $\kappa \simeq 1.52$ for the pulse of Eq.~(\ref{eq:Gausspls})
[$\kappa =1$ for the square, non-adiabatic pulse $\Om_\mathrm{t}(t) = \Om_\mathrm{t0}$].
In order to suppress the undesired phase shift, we assume that
the F\"orster defect $\delta \omega$ can be tuned to compensate
the level shift $\beta_{ab}$.
In the numerical simulations of Fig.~\ref{fig:errCZ} we thus
set $\delta \omega = - \beta_{ab}$. There we choose
$B_{rr(ab)} = B/2$ and $\delta \omega_{rr(ab)} = 3 B_{rr(ab)}$, but
other values of $B_{rr(ab)} < \delta \omega_{rr(ab)}$ yield similar
results for the gate fidelities under the adiabatic conditions.
For the non-adiabatic (square) pulse $\Omega_{\rt}$, however, we
observe significant population leakage to states $\ket{a_\rc' b_\rt'}$
and $\ket{b_\rc' a_\rt'}$, in addition to the non-vanishing residual
population of the bright states $\ket{\psi_\pm}$ after the pulse.
This explains the slightly larger (by a factor of $\sim 1.4$) gate
error for $B\tau \gtrsim 10^5$ as compared to the analytic estimates
which take into account only the population of the bright states.

In the numerical simulations of Fig.~\ref{fig:errCZ},
we neglect decay and decoherence of the qubit states $\ket{0_{\rc,\rt}}$
and $\ket{1_{\rc,\rt}}$ and assume that all the Rydberg states
$\ket{\rho_{\rc,\rt}}$ ($\rho = r,a,a',b,b'$) of atoms (c) and (t)
decay with the same rate $\Gamma = 1/\tau$. This process
is described by adding the Lindbladian decay term
$\mathcal{L}^2 = \sum_{j=\rc,\rt} \sum_{\rho} \Gamma \ket{\rho_{j}}\bra{\rho_{j}}$
to the Hamiltonian of Eq.~(\ref{eq:HamFull}), making it thus non-Hermitian,
\begin{equation}
\tilde{H} = H - \frac{i}{2} \hbar \mathcal{L}^2 . \label{eq:HamFullNH}
\end{equation}
We solve the Schr\"odinger equation
$\partial_t \ket{\Psi} = -\frac{i}{\hbar} \tilde{H}\ket{\Psi}$
for the total state-vector $\ket{\Psi}$ of the systems of two six-level
atoms. The decay of the Rydberg states thus results in the loss of
the total population of the system (decreasing the norm $\braket{\Psi}{\Psi}$
due to population of states outside the basis states). This slightly 
overestimates the gate error by disregarding processes that may re-populate 
the qubit states from the Rydberg states by the spontaneous decay.

In calculating the error probabilities for the phase gate,
for each two qubit input state 
$\ket{\Psi(t_{\mathrm{in}})} = \ket{00},\ket{01}, \ket{10},\ket{11}$,
we propagate the state-vector $\ket{\Psi(t)}$ of the system until
the end of the sequence ($t=t_{\mathrm{out}}$) involving the preparatory
microwave and optical pulses as described above. 
From the four output states $\ket{\Psi(t_{\mathrm{out}})}$, we obtain 
the $4\times 4$ transformation matrix $U$, and then calculate the average 
fidelity \cite{Pedersen2007} of the two-qubit gate via 
$F = [ \mathrm{Tr} (M \, M^{\dagger}) + |\mathrm{Tr}(M)|^2]/(20)$ with 
$M = U_{\textsc{cz}}^{\dagger} \, U$, where $U_{\textsc{cz}}$ is the transformation 
matrix of the ideal phase gate.  The average gate error is identified with 
the infidelity $E=1-F$.

\section{Rydberg blockade gate}
\label{ap:RBG}

For comparison, we now discuss the relevant properties of the Rydberg blockade 
gate performed in the conventional way \cite{Jaksch2000} via excitation
of (identical) Rydberg states $\ket{r_{\rc,\rt}}$ of the control and target atoms.

Strong interatomic interactions can be provided by either static or non-resonant
dipole-dipole interaction, which results in the energy shift of double Rydberg
excitation, $H_{\mathrm{sh}} = \hbar B_{\mathrm{sh}} \ket{r_\rc r_\rt} \bra{r_\rc r_\rt}$.
The static dipole-dipole interaction occurs between the Stark
eigenstates of the atoms in a static electric field.
Then atoms in such states $\ket{r}$ possess permanent dipole moments
$\wp_{r} \propto n^2$, leading again to $B_{\mathrm{sh}} = C_3/x^3$
with $C_3 \propto n^4$.
The dipole-dipole exchange interaction $B$ reduces to the van der Waals
type of interaction, $B_{\mathrm{sh}} = C_6/x^6$, with 
$C_6 \simeq C_3^2/\de \om_{\mathrm{F}} \propto n^{11}$, when the F\"orster defect
$\de \om_{\mathrm{F}} \propto n^{-3}$ between $\ket{r_\rc r_\rt}$ and (the nearest)
$\ket{a_\rc b_\rt}$ is large compared to $B$. More precisely, we
have to sum up the second order level shifts of $\ket{r_\rc r_\rt}$ due to
the non-resonant interaction with all the pairs of states $\ket{a_\rc b_\rt}$,
$C_6 \propto \sum_{ab} \frac{|\wp_{ar} \wp_{rb}|^2}{\om_r+\om_{r}-\om_{a}-\om_{b}}$.

%%%%%%%%%%%%%%%%FIGURE%%%%%%%%%%%%%%%%
\begin{figure}[t]
\includegraphics[width=5cm]{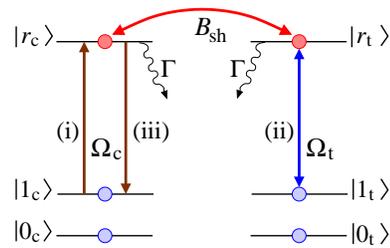}
\caption{(Color online)
Schematics of the conventional Rydberg-blockade phase gate
between the control and target atoms \cite{Jaksch2000}.
Atoms in Rydberg states $\ket{r_{\rc,\rt}}$ strongly interact with each other
via the static or non-resonant dipole-dipole interaction leading to
level shift $B_{\mathrm{sh}} (\gg \Om_\mathrm{t})$ of state $\ket{r_\rc r_\rt}$.}
\label{fig:alsRB}
\end{figure}
%%%%%%%%%%%%%%%%%%%%%%%%%%%%%%%%

We consider again the dynamics of the two-atom system
in state $\ket{r_\rc 1_\rt}$ during step (ii), but under the
scenario of dispersive Rydberg-Rydberg interaction leading to
large level shift $B_{\mathrm{sh}}$ of the double-excited state
$\ket{r_\rc r_\rt}$, see Fig.~\ref{fig:alsRB}.
The laser field with Rabi frequency $\Om_\mathrm{t}$ acts
on the transition $\ket{1} \to \ket{r}$ of the target atom,
which is now shifted out of resonance by the Rydberg-Rydberg
interaction $B_{\mathrm{sh}}$.
The Hamiltonian of the two-level system is then
\begin{equation}
H_{2}/\hbar = \hlf (\Om_\mathrm{t} \ket{r_\rc r_\rt}\bra{r_\rc 1_\rt} + \mathrm{H.c.})
+ B_{\mathrm{sh}} \ket{r_\rc r_\rt}\bra{r_\rc r_\rt}.  \label{eq:Ham2}
\end{equation}
The eigenstates of $H_{2}$ are
$\ket{\psi_{\pm}} = ( [\bar{\Om} \mp B_{\mathrm{sh}}] \ket{r_\rc 1_\rt}
\pm  \Om_{\mathrm{t}} \ket{r_\rc r_\rt})/\nu_{\pm}$, with
$\bar{\Om} \equiv \sqrt{B_{\mathrm{sh}}^2 + \Om_{\mathrm{t}}^2}$ and
$\nu_{\pm}^2 = 2 (B_{\mathrm{sh}}^2 \mp \bar{\Om} B_{\mathrm{sh}}
+ \Om_{\mathrm{t}}^2)$, and the corresponding eigenvalues are
$\la_{\pm} = \hlf (B_{\mathrm{sh}} \pm \bar{\Om})$.

\subsection{Adiabatic driving}

For a smooth $2\pi$-pulse $\Om_{\mathrm{t}}(t)$, the initial state $\ket{r_\rc 1_\rt}$
is adiabatically connected to the $\ket{\psi_{-}}$ eigenstate with
eigenvalue $\la_-$, and non-adiabatic transition to $\ket{\psi_{+}}$
is suppressed for $\Om_{\mathrm{t}0} \ll B_{\mathrm{ex}}$. During the pulse,
the population of the double-excited Rydberg state $\ket{r_\rc r_\rt}$
is $P_{\mathrm{Ry}} \simeq \frac{\Om_t^2}{4 B_{\mathrm{ex}}^2}$,
which returns back to $\ket{r_\rc 1_\rt}$ at the end of the pulse,
$\Om_\mathrm{t}(t=T_\mathrm{t}) \to 0$.

The situation is thus similar to that of the resonant exchange interaction
studied in the main text, but there are also important differences.
Since the adiabatically connected eigenstate $\ket{\psi_-}$ has non-zero
energy $\la_-(t) \simeq - \frac{\Om_{\mathrm{t}}^2(t)}{4 B_{\mathrm{sh}}}$,
at the end of the $\Om_{\mathrm{t}}(t)$ pulse, state $\ket{r_\rc 1_\rt}$
acquires the phase $\phi = \int_0^{T_{\mathrm{t}}} \la_-(t) dt$.
If both $\Om_{\mathrm{t}}$ and $B_{\mathrm{sh}}$ are well-defined,
this phase is known and can be amended, as described in \cite{Zhang2012}.
[In the present context, correcting the phase shift $\phi$ involves
splitting during step (ii) the $2\pi$ pulse $\Om_{\mathrm{t}}(t)$ into
two smooth $\pi$-pulses $\Om_{\mathrm{t}}^{(1)}(t)$ and
$e^{i \phi} \Om_{\mathrm{t}}^{(2)}(t)$ with the relative phase difference $\phi$,
and then, after step (iii), applying to the target qubit the operation
$\hat{Z}_{\mathrm{t}} = \ket{0_\rt}\bra{0_\rt} + e^{-i \phi} \ket{1_\rt}\bra{1_\rt}$.]
But if there is an uncertainty in the interaction strength, 
$B_{\mathrm{sh}} \to B_{\mathrm{sh}} + \de B$ with 
$\de B \simeq \partial_x B_{\mathrm{sh}}|_{x=x_0} \de x$,
due to uncertainty $\de x$ in the interatomic distance $x_0$, 
it will cause phase errors of the target qubit,
$\de \phi \simeq \pi \Om_{\mathrm{t}} \de B/B_{\mathrm{sh}}^2$.
Furthermore, during the pulse $\Om_{\mathrm{t}}$ the pair of atoms
in state $\ket{\psi_-}$ occupy the double Rydberg excitation state
$\ket{r_\rc r_\rt}$ with a finite probability $P_{\mathrm{Ry}}$
and, hence, experience a mechanical
force $F = - \hbar \partial_x \la_- |_{x=x_0}$, where the $x$ dependence
of $\la_-$ stems from $B_{\mathrm{sh}}$.

\subsection{Square pulse of specific amplitude}

Perhaps surprisingly, the rotation errors can in principle be avoided
even when using non-adiabatic, square $2\pi$-pulses $\Om_{\mathrm{t}}$
on the target atom, as was studied in detail in Ref. \cite{Shi2017}. 
The time-dependent state of the two-level system described by 
Hamiltonian~(\ref{eq:Ham2}) can be written as 
$\ket{\psi(t)} = c_1(t) \ket{r_\rc 1_\rt} + c_r(t) \ket{r_\rc r_\rt}$, with
$c_1(0) = 1$ and $c_r(0) = 0$ corresponding to $\ket{\psi(0)} = \ket{r_\rc 1_\rt}$.
Neglecting the decay $\Ga \ll \Om_{\mathrm{t}}$, the general solution for
the amplitudes of state-vector $\ket{\psi(t)}$ is given by \cite{PLDP2007}
\begin{subequations}
\begin{eqnarray}
c_1(t) &=&  e^{i \phi(t)} \left[\cos(\hlf \bar{\Om} t)  -
i \frac{B_{\mathrm{sh}}}{\bar{\Om}} \sin(\hlf \bar{\Om} t) \right] , \\
c_r(t) &=& e^{- i \phi(t) - \phi_{\mathrm{t}}} \frac{\Om_{\mathrm{t}}}{\bar{\Om}}
\sin(\hlf \bar{\Om} t) ,
\end{eqnarray}
\end{subequations}
where $\phi(t) \equiv \hlf B_{\mathrm{sh}} t$ and $\phi_{\mathrm{t}}$ is
the laser phase. Our goal is that $c_r(t = T_{\mathrm{t}}) = 0$
at time $T_{\mathrm{t}} = 2\pi/\Om_{\mathrm{t}}$ of the resonant $2\pi$ pulse.
We thus require that $\hlf \bar{\Om} T_{\mathrm{t}} = 2\pi k$
($k \in \mathbf{N}$). We obtain
$\Om_{\mathrm{t}} = \frac{B_{\mathrm{sh}}}{\sqrt{4k^2 - 1}}$, which is largest
for $k = 1$: $\Om_{\mathrm{t}} = \frac{1}{\sqrt{3}} B_{\mathrm{sh}}$.
The final phase of state $\ket{r_\rc 1_\rt}$ would then be $\phi = \sqrt{3} \pi$,
which should be amended as described above. Notice, however, that if there
are pulse timing/amplitude errors and/or uncertainly in $B_{\mathrm{sh}}$,
the averaged (over a small time interval $\De t \simeq \pi/\bar{\Om}$)
residual Rydberg population of the target atom will be
$P_{\mathrm{Ry}} \simeq \frac{\Om_t^2}{2 B_{\mathrm{sh}}^2}$.
Hence, this method is even less robust with respect to uncertainties
of parameters than the adiabatic methods above.

\subsection{Gate error estimates}

%%%%%%%%%%%%%% TABLE %%%%%%%%%%%%%%%%%%
\begin{table}[t]
\caption{\label{tab:errors}%
Error probabilities during step (ii) of the Rydberg-blockade gate,
for four two-qubit input states.}
\begin{ruledtabular}
\begin{tabular}{lccc}
\textrm{Input state}&
\textrm{Decay error\footnote{For the control and/or target qubit in state $\ket{1}$}}&
\textrm{Rotation error\footnote{Only for non-adiabatic (square) pulse $\Om_{\mathrm{t}}$}}&
\textrm{Phase error}\\
\colrule
$\ket{0_\rc 0_\rt}$ & 0 & 0 & 0\\
$\ket{0_\rc 1_\rt}$ & $\frac{\pi \Ga}{\Om_{\mathrm{t}}}$ & 0 & 0\\
$\ket{1_\rc 0_\rt}$ & $\frac{2\pi \Ga}{\Om_{\mathrm{t}}}$ & 0 & 0\\
$\ket{1_\rc 1_\rt}$ & $\frac{2\pi \Ga}{\Om_{\mathrm{t}}}
+ \frac{\pi \Ga \Om_{\mathrm{t}}}{4 B^2}$
& $\frac{\Om_{\mathrm{t}}^2}{2 B^2}$
& $\frac{\pi \de B \, \Om_{\mathrm{t}}}{B^2}$\\
\end{tabular}
\end{ruledtabular}
\end{table}
%%%%%%%%%%%%%%%%%%%%%%%%%%%%%%%%%%%%

Let us summarize the above results.
In Table~\ref{tab:errors} we show the error probabilities during
step (ii) of the blockade gate, for the four two-qubit
input states. For the gate performed with a smooth (adiabatic) pulse
$\Om_{\mathrm{t}}$, and assuming compensation of the interaction phase $\phi$,
the error averaged over all the inputs is again
$E \simeq \frac{\pi \Ga}{4} \left[\frac{5}{\Om_{\mathrm{t}}}
+ \frac{\Om_{\mathrm{t}}}{4 B^2}\right]$, as in the main text
for the dark-resonance gate.
Choosing $\Om_{\mathrm{t}} = \alpha B$ with $\alpha \ll 1$, we obtain
$E \simeq \eta \frac{\Ga}{B}$ with $\eta \simeq \frac{5 \pi}{4 \alpha}$.
If we also include the phase error of the gate due to uncertainty
in the dispersive Rydberg-Rydberg interaction strength,
this coefficient would increase accordingly,
$\eta \simeq \frac{5\pi}{4 \alpha} + \frac{\pi \alpha}{4} \frac{\de B}{\Ga}$.

\end{document}